\documentclass[a4paper,11pt]{article}
\usepackage[T1,T2A]{fontenc}
\usepackage[utf8]{inputenc}
\usepackage{amsmath,amsthm,amssymb}
\usepackage{graphics}
\usepackage{graphicx}
\begin{document}
\title{Relativistic spin-Hall effect in an external magnetic field in Al and Pt.}
\author{Yu.N. Chiang (Tszyan), M.O. Dzyuba\\
\emph{B. I. Verkin Institute for Low Temperature Physics and Engineering,}\\
\emph{National Academy of Sciences of Ukraine.}\\
\emph{Nauky Ave. 47, Kharkov 61103, Ukraine}\\
\emph{E-mail:chiang@ilt.kharkov.ua}}
\date{}
\maketitle
\begin{abstract}

The relativistic Rashba contribution to the spin-Hall effect in external electric and magnetic fields in Al and Pt was investigated. Schemes of edge accumulation of spins are proposed that take into account the flip of spins when they do not coincide with the direction of the magnetic field. Based on the obtained experimental data on the spin-Hall effect, an assessment of the spin-orbit interaction in the studied paramagnets was made.
\end{abstract}

As is well known, spin-orbit interaction (SO) is the interaction of some effective magnetic field $\emph{\textbf{B}}_{i}$, arising in the coordinate system of a moving electron, with its magnetic moment-spin. The Hamiltonian of such interaction, following from the complete Dirac equation of motion as a correction of the order of $v^{2}/c^{2}$ (\emph{c} is the speed of light), is relativistic. The effects described by such a relativistic Hamiltonian also include the Rashba effect [1, 2], which is a spin splitting phenomenon arising due to spin-orbit interaction in electron systems without out-of-plane mirror symmetry. The proper effective magnetic field, caused by the spin-orbit interaction, is rigidly related to the momentum $\emph{\textbf{p}}$ and the electron mass $\emph{m}$, characterizing a typical relativistic effect in a solid induced by the spin splitting of energy bands at wave vectors $\emph{\textbf{k}}$ that differ from the time-reversal-invariant moments:
\begin{equation}\label{1}
\emph{\textbf{B}}_{\emph{i}} \sim \mathcal{\emph{\textbf{E}}}\times \emph{\textbf{p}}/\emph{m}\emph{c}^{2}.
\end{equation}
 
 If the electron is in an electric field $\emph{\textbf{E}}$ that removes the center of spatial inversion, then the Rashba SO contribution to the effective Hamiltonian, induced by the spin splitting of the energy bands,  can be
   \begin{equation}\label{2}
   \mathcal{H}_{\rm SO}=
   \beta e[\hat{\sigma}\times \emph{\textbf{k}}] \emph{\textbf{E}},
    \end{equation}
leading to the removal of the double spin degeneracy of the energy bands (Fig. 1).~ ($\beta$ is the band coefficient determined by the point symmetry of the crystal, $\hat{\sigma}$ is a pseudovector composed of the Pauli matrices)

\begin{figure}
  \centering
  \includegraphics[width=7cm]{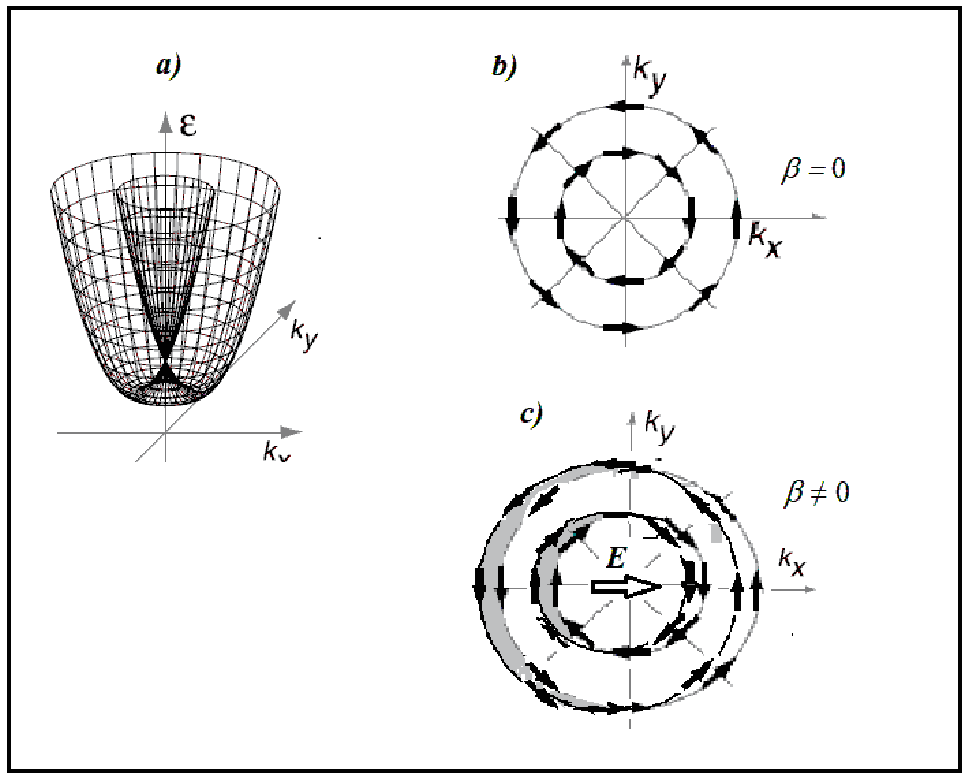}
  \caption{(\emph{a}) Splitting of doubly degenerate energy levels at non-centrosymmetric points of the Brillouin zone due to the SO interaction; (\emph{b}) spin orientation in eigenstates in the case of the Rashba effect; (\emph{c}) the same in an external electric field $\textbf{E}$, causing edge accumulation of spins.}\label{1}
  \end{figure}

Here we consider the effects in the conduction band of aluminum and platinum, where spin-orbit coupling arises due to $\emph{\textbf{k}}\cdot\emph{\textbf{p}}$ hybridization, in particular, of the \emph{s} and \emph{p} bands in Al and the \emph{s} and \emph{d} bands) in Pt. The parameters that determine the effective magnetic field $\emph{\textbf{B}}_{i}$ in Al and Pt samples were (in CGSE units): $E \sim 7.5\cdot 10^{-3} cm^{-1/2}\cdot g\cdot s^{-1}(~2.5\cdot 10^{-7}{\rm V}/m); ~p_{\rm F} \sim 2\cdot 10^{-19}cm\cdot g\cdot s^{-1};~ mc^{2}\approx 10^{-8}g\cdot cm^{2}\cdot s^{-2}$. Thus, the intrinsic relativistic magnetic field in both metals can be $B_{i}\sim10^{-13}\rm G$. In the absence of external magnetic fields in the presence of an electric field $\emph{\textbf{E}}$, the field $B_{i}$ allows the spin-Hall effect (SHE) to manifest itself, which was reported previously [3, 4, 5]. Moreover, the given value of $B_{i}$ indicates the possibility of manipulating the directions of spin polarization and, consequently, spin currents using external magnetic fields practically any reasonable magnitude, which is of great importance for spintronics.
\begin{figure}[h!]
  \centering
  \includegraphics[width=15cm]{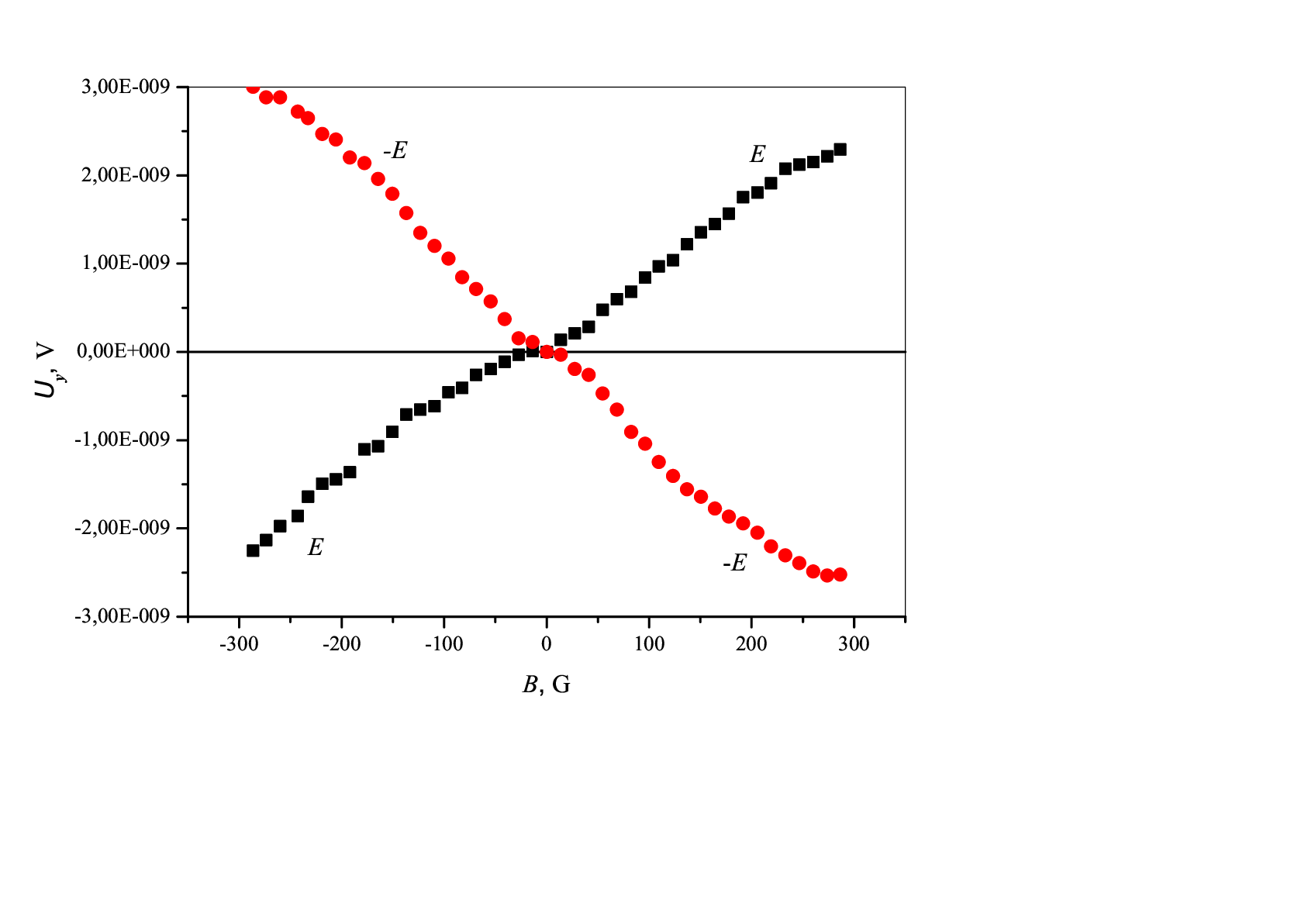}
  \vspace{-2cm}
  \caption{Voltage $U_{y}$ of an asymmetric aluminum sample, measured as a function of external magnetic $B^{\pm z}$ and d.c. electric $E^{\pm x}$ fields at \emph{T} = 4.2 K. (Sample geometry: $L_{x}=1cm; L_{y}=5mm; (A_{xz}^{1}/A_{xz}^{3})\approx 5; ~$electron mean free path $l_{c}\sim 0.15 mm;~ L_{y} > \lambda_{sf}$.)}\label{2}
\end{figure}

In this paper, we rigorously justify the procedure for extracting the contribution of the nanovolt-level SHE from the total value of the signal $U_{y}$ measured in an external magnetic field oriented along the \emph{z} axis, at a direct current $\mathbf{\textbf{I}}$ along \emph{x}, including the contribution of the standard Hall effect (CHE), which is several orders of magnitude greater than the contribution of SHE (Fig. 2):
\begin{equation}\label{3}
   \mathbf{U}_{y} = \mathbf{U}_{\rm CHE} + \mathbf{U}_{\rm SHE},
\end{equation}
where $ \mathbf{U}_{\rm CHE} = (R_{\rm CHE}/R_{L})(L_{x}/d_{z})[\mathbf{B}^{\pm z}_{ext} \times \mathbf{E}^{\pm x}]$ ($R_{\rm CHE}$ is the Hall constant; $L_{x}$ is the sample length along \emph{x}; $R_{L}$ is the sample resistance at the length $L_{x}; d_{z}$ is the sample size along \emph{z}; $\mathbf{B}_{ext}$ is the external magnetic field along \emph{z}; $\mathbf{E}^{\pm x}$ is the electric field strength (see Fig. 3, 4). As for the contribution of CHE to the self-magnetic field $\emph{\textbf{B}}_{i}$, then it is essentially of the same order as Rashba's contribution, and in this case (Fig. 3) we will not take it into account.
\begin{figure}[htb]
  \centering
  \includegraphics[width=10cm]{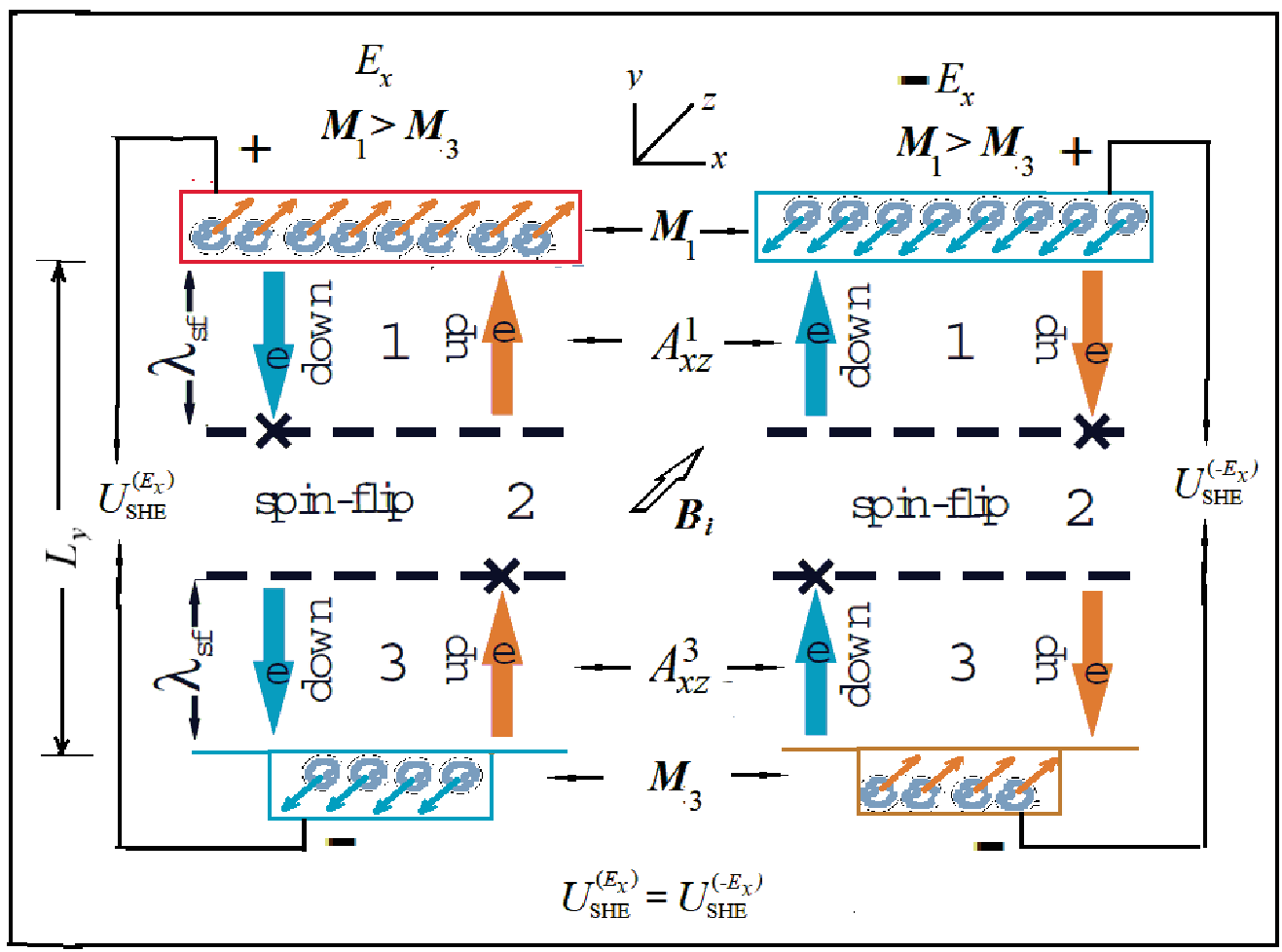}
  \caption{Scheme of proper edge accumulation of spins in asymmetric samples of normal metal with a spin relaxation region (2) of length [$L_{y} - 2\lambda_{sf}$] ~ ($\emph{\textbf{B}}_{i}$ is the intrinsic magnetic field of the metal, $L_{y}$ is the size of the sample along \emph{y}).}\label{3}
\end{figure}

It is clear that eliminating the contribution of $\mathbf{U}_{\rm CHE}$ from $\mathbf{U}_{y}$ under conditions of a large external magnetic field requires at least two measurements with switching of the directions of either $\mathbf{B}_{ext}$ or $\mathbf{E}$. Preference should be given to changing the sign of $\mathbf{E}$ (achieved by changing the sign of the direct current $\mathbf{I}\|\mathbf{E}$), in view of the technical possibility of stabilizing and reproducing the current value with an accuracy of $\triangle I/I \approx 10^{-5}$, exceeding the assumed resolution of $\mathbf{U}_{\rm SHE}$ at the level of $(\mathbf{U}_{\rm SHE}/\mathbf{U}_{\rm CHE}) \sim 10^{-3}$, while commutation of the direction of the magnetic field $\mathbf{B}_{ext}$ due to hysteresis phenomena can introduce an error no less than $\mathbf{U}_{\rm SHE}$.

In Figure 1, the orientation of spins in eigenstates (\emph{b}) and under the application of an electric field (\emph{c}) is shown for the relativistic Rashba effect (in the absence of an external magnetic field), and in Fig. 3, the scheme of the edge accumulation of spins in paramagnets with one type of charge carriers (Al, Pt, [5, 6]) for the case of \emph{c} from Fig. 1 according to the theory of Dyakonov and Perel [7] is shown. In Fig. 4, a scheme of such accumulation in an external magnetic field is shown, \emph{for the first time} taking into account the influence of the latter on the orientation of spins, namely, assuming a spin flip for mismatched relative orientations of the field and spin (this effect does not apply to the case of $\emph{\textbf{B}}_{i}$).
\begin{figure}[htb]
  \centering
  \includegraphics[width=15cm]{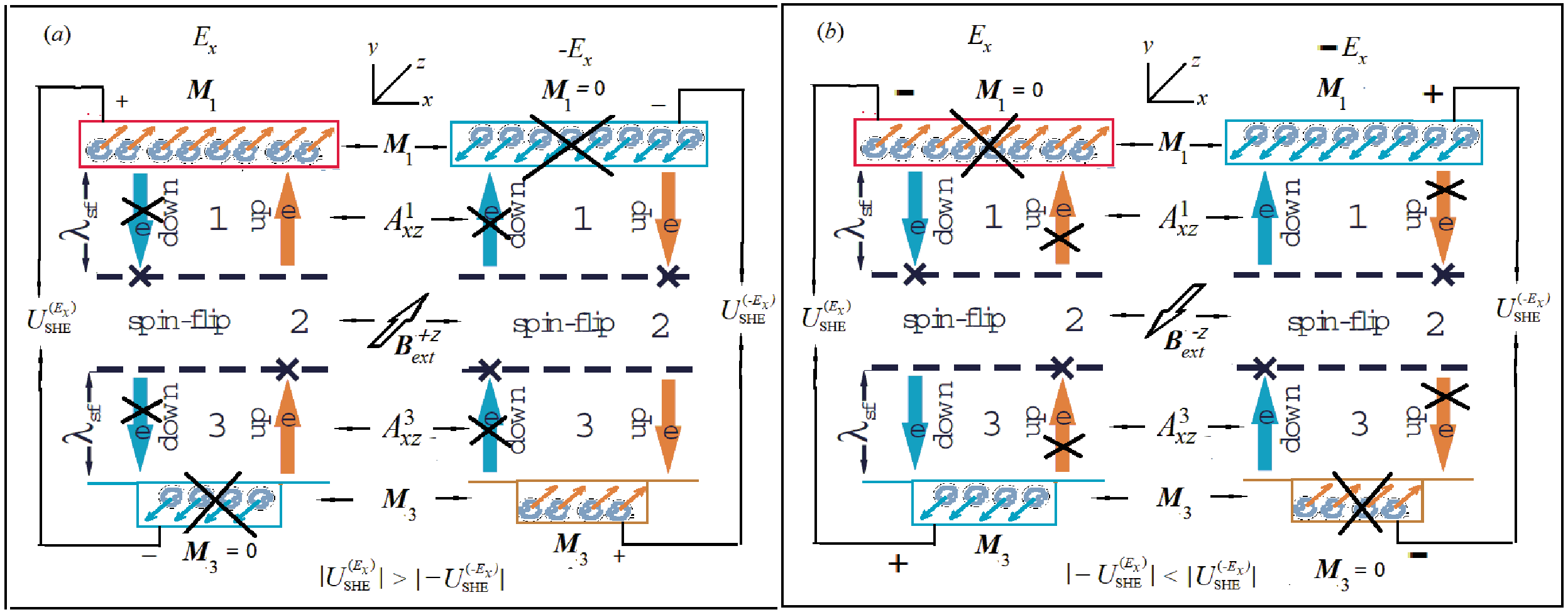}
  \caption{Schemes of edge accumulation of spins in external magnetic $\mathbf{B}_{ext}$ and electric $\textbf{E}$ fields. (\emph{a}) Flip of spins "down", polarized against "$B_{z}$"; (\emph{b}) Flip of spins "up", polarized against "$-B_{z}$".}\label{4}
\end{figure}

As we have repeatedly emphasized [5, 6], the occurrence of a transverse nonequilibrium spin-charge imbalance as a spin-Hall effect under conditions in which the sample dimensions, at least in the \emph{y} direction, do not exceed the spin relaxation length would be impossible. Since the dynamics of spins of opposite orientation are opposite in sign and equally probable, in this case the differently directed spin currents across the width of the sample would be continuous and equal, creating an equilibrium accumulation. In [5], we proposed an asymmetric geometry of samples, symbolically depicted in Figures 3 and 4, providing a nonequilibrium distribution of spins along the edges of the samples with the following relationship of microscopic and dimensional parameters: $L_{y} > \lambda_{sf}$ and $A_{xz}^{1} > A_{xz}^{3}$, where $L_{y}$ is the sample width, $\lambda_{sf}$ is the spin relaxation length, $A_{xz}$ are the cross-sections of regions 1 and 3 in the \emph{y} direction, and region 2 is the region of spin depolarization. In this regard, the effect considered here can be called the geometric spin-Hall effect (GSHE).

Figure 3 shows the proposed scheme of edge accumulation of spins in paramagnetic samples (Al, Pl) with similar geometry in the regime of the intrinsic magnetic field $\emph{\textbf{B}}_{i}$ interacting with an external electric field in two directions of the latter. Taking into account the destruction of spin currents beyond the spin relaxation limits, we obtain a spin-charge imbalance with a voltage $U_{y}$ reflecting the relativistic Rashba contribution to the spin-Hall effect. From this accumulation scheme follows the independence of the sign of the spin-charge voltage from the direction of $\textbf{E}$, along with a change in the sign of the spin magnetization $|\mathbf{\emph{\textbf{M}}}_{1} - \mathbf{\emph{\textbf{M}}}_{3}|$:
\begin{equation}\label{4}
  U^{(+E_{x})}_{\rm SHE} = U^{(-E_{x})}_{\rm SHE}; ~|\mathbf{\emph{\textbf{M}}}_{1} - \mathbf{\emph{\textbf{M}}}_{3}|^{+E_{x}} = - |\mathbf{\emph{\textbf{M}}}_{1} - \mathbf{\emph{\textbf{M}}}_{3}|^{-E_{x}}
\end{equation}

The magnitude of the measured voltage  at two current values, 1A and 2A,  in two directions of the magnetic field in Al and Pt was a very small value of the order of $(10^{-11} \div 10^{-10})$V.

\begin{figure}[ht]
  \centering
  \includegraphics[width=15cm]{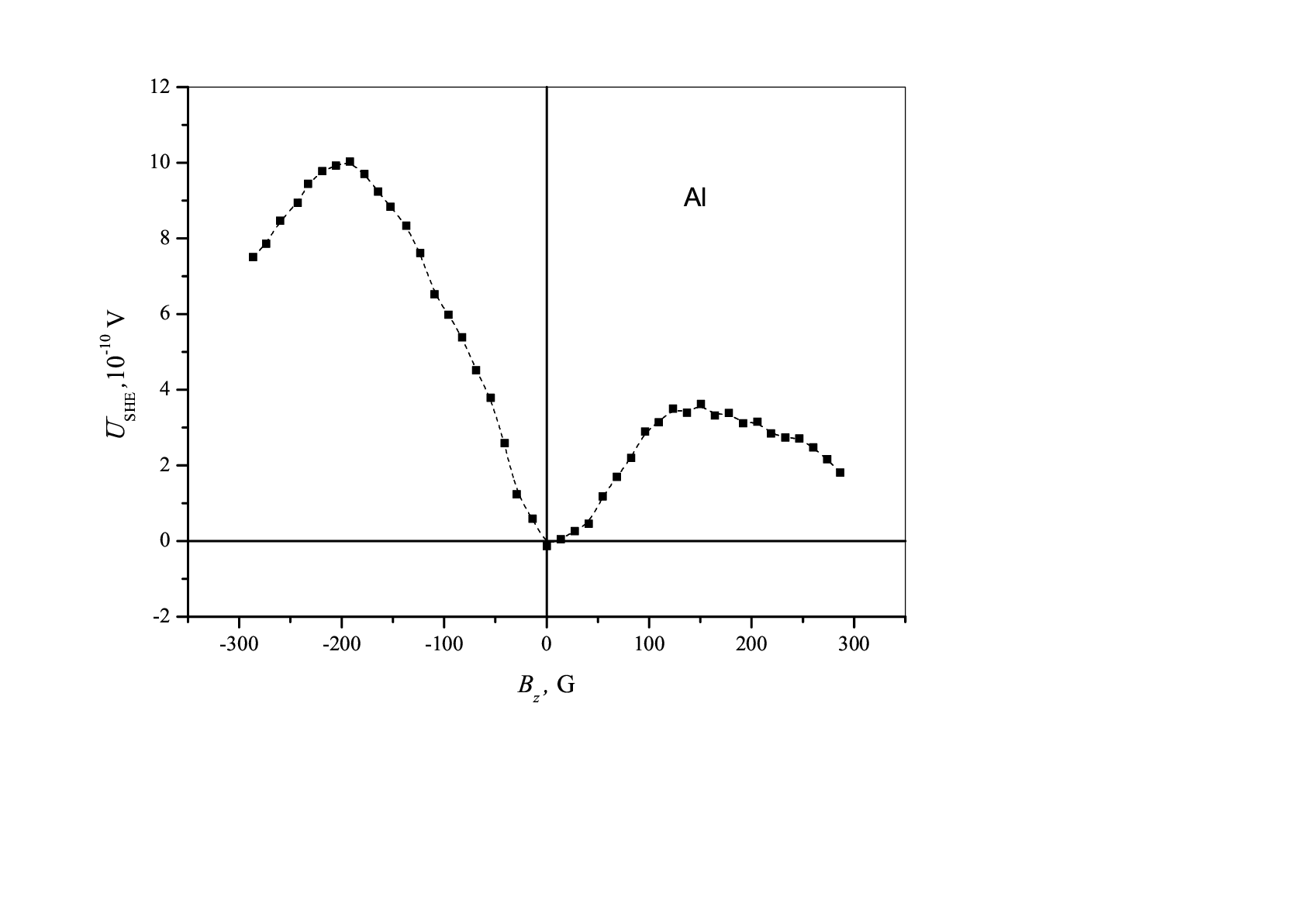}\vspace{-2cm}
  \caption{Dependence of the relativistic spin-Hall effect in aluminum on the magnitude of the external magnetic field (in a sample with the geometry shown in Fig. 2).}\label{5}
\end{figure}

Figure 4 shows the schemes of the edge accumulation of spins in external magnetic $\mathbf{B}_{ext}$ and electric $\textbf{E}$ fields taking into account both the discontinuities of spin currents at the boundaries of space of the order of $\lambda_{sf} < L_{y}$ and the flip of spins when their orientation does not coincide with the direction of $\mathbf{B}_{ext}$. As was justified above, to calculate $U_{\rm SHE}$, pairs of measurements were used at two opposite directions of the electric field with a fixed direction of the external magnetic field, which made the excluded contribution $U_{\rm CHE}$ symmetrical up to $\frac{\Delta E^{\pm}_{x}}{E^{\pm}_{x}}\ll \frac{U_{\rm SHE}}{U_{\rm CHE}}$. As a result, the averaged spin-charge SHE for two directions of $\mathbf{B}_{ext}$ in accordance with the accumulation schemes in this figure will be equal
\begin{equation}\label{5}
  \begin{aligned}
  \overline{U}_{\rm SHE}^{+B_{z}} = \frac{1}{2}[U_{\rm SHE}^{+E_{x}} - U_{\rm SHE}^{-E_{x}}]^{+B_{z}} = \frac{1}{2}[U_{y}^{+E_{x}} + U_{y}^{-E_{x}}]^{+B_{z}}; \\
    \overline{U}_{\rm SHE}^{-B_{z}} = \frac{1}{2}[U_{\rm SHE}^{-E_{x}} - U_{\rm SHE}^{+E_{x}}]^{-B_{z}} = \frac{1}{2}[U_{y}^{+E_{x}} + U_{y}^{-E_{x}}]^{-B_{z}}.
\end{aligned}
\end{equation}
  
  The corresponding curves of the dependences of the relativistic spin-Hall effect in aluminum and platinum on the magnitude of the external magnetic field, obtained from expressions (5), are shown in Figs. 5 and 6. We see that the nature of the curves is qualitatively different from the almost linear dependence of the contribution of $U_{\rm CHE}$ on the magnetic field (see Fig. 2 and [5]) and is consistent with the close-to-parabolic behavior discussed in [6]. The obtained values of the spin-Hall effect in both metals make it possible to obtain the real ratio of the spin-orbit relativistic energy shifts in these metals, which at $B_{z} = \rm 200 G$ according to expression (2) of the order

 \begin{equation}\label{6}
  \frac{ \beta^{Pt}}{\beta^{Al}} \geq \frac{ \mathcal{H}_{\rm SO}^{Pt}}{ \mathcal{H}_{\rm SO}^{Al}}\cdot \frac{\sigma_{y}^{Al}}{\sigma_{y}^{Pt}} = \frac{U_{\rm SHE}^{Pt}}{U_{\rm SHE}^{Al}}\cdot \frac{\sigma_{y}^{Al}}{\sigma_{y}^{Pt}} \sim 40\times 3 = 120.
 \end{equation}

\begin{figure}[h]
  \centering
  \includegraphics[width=12cm]{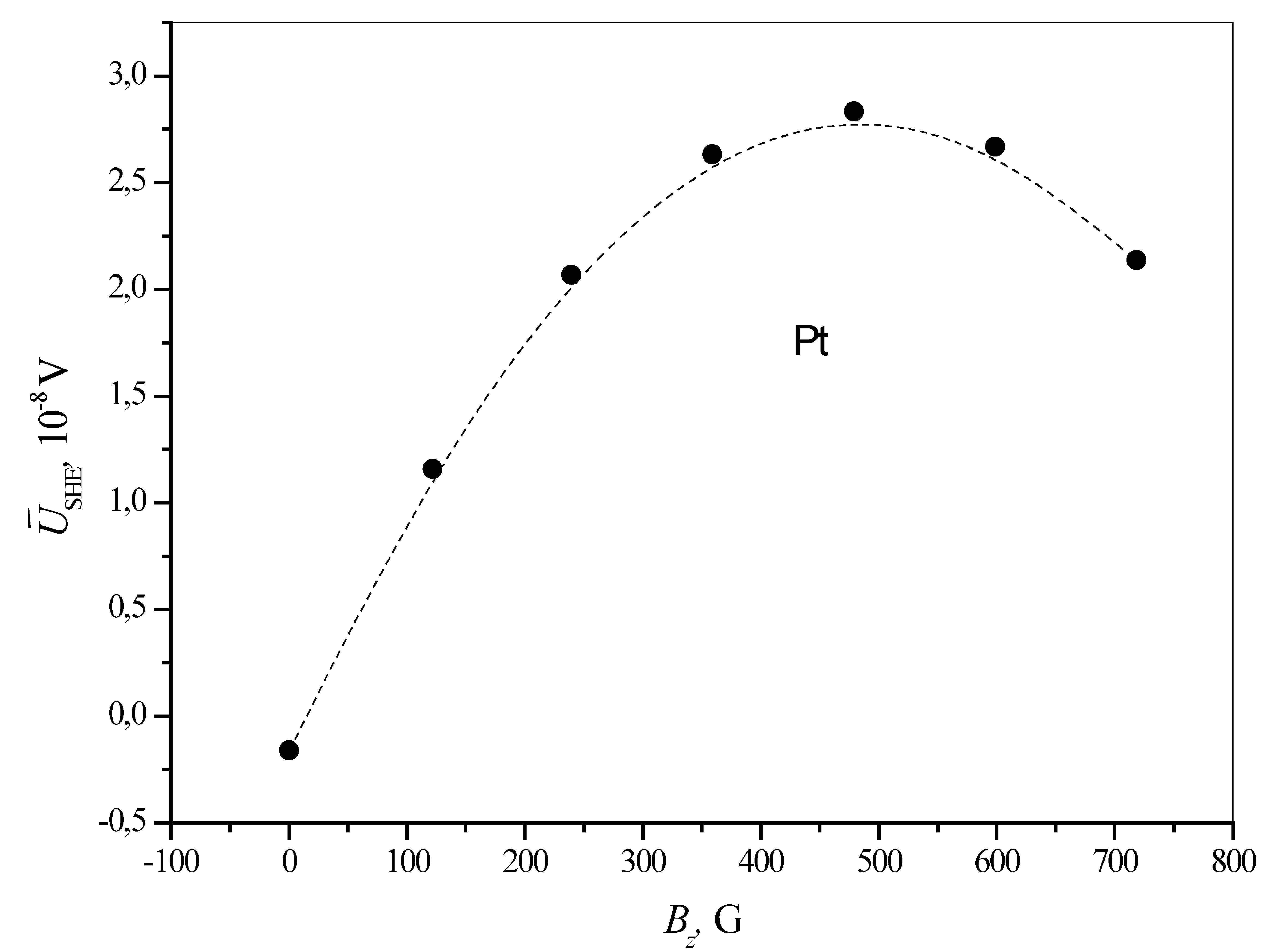}
  \caption{Dependence of the relativistic spin-Hall effect in platinum on the external magnetic field at \emph{T} = 13 K. (Sample geometry: $L_{x}= 0.5 cm; L_{y}=2 mm; (A_{xz}^{1}/A_{xz}^{3})\approx 10; ~$electron mean free path $l_{c}\sim 0.1 mm;~ L_{y} > \lambda_{sf}$.)}\label{6}
\end{figure}

This difference in the Rashba contributions directly indicates a stronger spin-orbit interaction in Pt than in Al, leading to spin-orbit splitting of different bands (in the \emph{p}-band for Al and in the \emph{d}-band for Pt.

In conclusion, we investigated the behavior of the relativistic spin-Hall effect in Al and Pt paramagnets in intrinsic and external magnetic fields, obtaining evidence of spin polarization in the direction of the external field. Only those measurements were considered in which the contribution of the conventional Hall effect during commutation could be considered completely symmetric in magnitude with an accuracy no less than the value of the sought SHE. A comparison of the spin-orbit interaction in aluminum and platinum was performed. All measurements were performed using a superconducting zero-picovoltmeter (SNPV) [8, 9].
 
\end{document}